\documentclass[a4paper,11pt]{article}
\pdfoutput=1 

\usepackage{jheppub2} 
\usepackage{bbm,bm,graphicx,mathtools,color,slashed,hyperref}
\usepackage{amssymb,amsmath}

\usepackage[compat=1.1.0]{tikz-feynhand}
\usetikzlibrary{patterns}
\usetikzlibrary{decorations.markings}

\graphicspath{{./Figures/}}

\newcommand{\diff}{\mathrm{d}}
\newcommand{\p}{\partial}
\newcommand{\ve}{\varepsilon}
\newcommand{\Diff}{{\mathcal{D}}}

\newcommand{\tr}{\mathrm{tr}}
\newcommand{\im}{\mathrm{i}}

\newcommand{\calD}{\mathcal{D}}

\newcommand{\calZ}{\mathcal{Z}}

\newcommand{\rme}{\mathrm{e}}

\DeclareMathOperator{\diag}{diag}

\preprint{KYUSHU-HET-258, OU-HET-1177, YITP-23-37}

\title{Topology of $SU(N)$ lattice gauge theories coupled with $\mathbb{Z}_N$ $2$-form gauge fields}

\author[1]{Motokazu Abe,}
\affiliation[1]{Department of Physics, Kyushu University, 744 Motooka,
Nishi-ku, Fukuoka 819-0395, Japan}

\author[2]{Okuto Morikawa,}
\affiliation[2]{Department of Physics, Osaka University, Toyonaka, Osaka
560-0043, Japan}

\author[1]{Soma Onoda,}

\author[1]{Hiroshi Suzuki,}

\author[3]{Yuya Tanizaki}
\affiliation[3]{Yukawa Institute for Theoretical Physics, Kyoto University,
Kitashirakawa Oiwakecho, Sakyo-ku, Kyoto 606-8502, Japan}
\emailAdd{yuya.tanizaki@yukawa.kyoto-u.ac.jp}

\abstract{
We extend the definition of L\"uscher's lattice topological charge to the case
of $4$d $SU(N)$ gauge fields coupled with $\mathbb{Z}_N$ $2$-form gauge fields.
This result is achieved while maintaining the locality, the $SU(N)$ gauge
invariance, and $\mathbb{Z}_N$ $1$-form gauge invariance, and we find that the
manifest $1$-form gauge invariance plays the central role in our construction.
This result gives the lattice regularized derivation of the mixed 't~Hooft
anomaly in pure $SU(N)$ Yang--Mills theory between its $\mathbb{Z}_N$ $1$-form
symmetry and the $\theta$ periodicity.
}

\begin{document}
\maketitle


\section{Introduction}
\label{sec:intro}

Gauge fields in the continuum description enjoy topological classifications,
and those in different topological sectors cannot be continuously deformed from
one another. This fact allows us to introduce a new parameter, called the
$\theta$ angle, in quantum Yang--Mills theories~\cite{Belavin:1975fg,%
Callan:1976je,Jackiw:1976pf}, and the presence of topological sectors makes the
dynamics of Yang--Mills theories quite rich and highly nontrivial. However, we
should note that it is nontrivial if these considerations based on topology are
robust under quantum fluctuations. Quantum field theories (QFTs) are subject to
ultraviolet divergences, and we need to introduce some regularization to
control it. Lattice regularization achieves it in a gauge invariant manner, but
it becomes unclear if the notion of continuous gauge fields survives under this
process.

L\"uscher addressed this issue for $SU(2)$ gauge theories and pointed out the
presence of topological sectors on the lattice by introducing the admissibility
condition~\cite{Luscher:1981zq} (see also Ref.~\cite{Phillips:1986qd}). There,
the topological charge on the lattice is explicitly defined in a local and
gauge-invariant way, and one can extend the construction to simple and simply-connected
gauge groups such as $SU(N)$ straightforwardly. In this paper, we consider its
extension for non-simply-connected gauge groups, in particular
$SU(N)/\mathbb{Z}_N$.

$SU(N)/\mathbb{Z}_N$ gauge fields can be thought of the $SU(N)$ gauge theories
coupled to $\mathbb{Z}_N$ $2$-form gauge fields. Treating the $\mathbb{Z}_N$
$2$-form gauge fields as the background field, it is equivalent to the $SU(N)$
gauge theories in the 't~Hooft twisted boundary
condition~\cite{tHooft:1979rtg}. van~Baal computed the topological charge for
smooth gauge fields in this setup, and it turns out that the topological charge
has a fractional shift in the unit of~$1/N$~\cite{vanBaal:1982ag}. These
observations now acquire renewed interest from the viewpoint of generalized
symmetries in QFTs. Pure $SU(N)$ Yang--Mills theory enjoys the $\mathbb{Z}_N$
$1$-form symmetry as the global symmetry~\cite{Gaiotto:2014kfa}, and the
$\mathbb{Z}_N$ $2$-form gauge field obtains a natural interpretation as the
gauge field coupled to the $1$-form
symmetry~\cite{Kapustin:2013qsa,Kapustin:2014gua}. Studying the response of
background gauge fields provides a concise and systematic way to extract
nontrivial consequences out of global symmetries, and the fractional shift of
the topological charge turns out to have a huge impact on our understanding of
the Yang--Mills vacua~\cite{Gaiotto:2017yup}.

In this paper, we construct the topological charge of lattice
$SU(N)/\mathbb{Z}_N$ gauge fields that maintains the locality, $SU(N)$ gauge
invariance, and $\mathbb{Z}_N$ $1$-form gauge invariance. Introducing the
admissibility condition for $SU(N)$ gauge fields coupled to flat $\mathbb{Z}_N$
$2$-form gauge fields, we show that the $SU(N)$-valued transition functions can
be locally defined from the lattice $SU(N)/\mathbb{Z}_N$ gauge fields. We show
that those transition functions satisfy the cocycle condition up to the center
elements specified locally by the $2$-form gauge fields. Thus, the principal
$SU(N)/\mathbb{Z}_N$ bundle is obtained from the admissible lattice gauge
fields, and we can compute their topological charge as its second Chern class.
We note that all of this procedure is completely parallel to L\"uscher's one
in~Ref.~\cite{Luscher:1981zq}. It turns out that we can keep the manifest
$1$-form gauge invariance at every stage of our construction, which is the
essential ingredient to circumvent various complications potentially caused by
the presence of higher-form gauge fields. Using this lattice topological
charge, we define the lattice Yang--Mills theory with the $\theta$~angle
coupled to the background $\mathbb{Z}_N$ $2$-form gauge fields, and we obtain
the anomalous relation for the Yang--Mills partition function on the lattice.
Our result justifies the observation of~Ref.~\cite{Gaiotto:2017yup} that had
been obtained in the continuum analysis assuming smoothness of gauge fields, so
we rigorously realize the mixed 't~Hooft anomaly between the $\mathbb{Z}_N$
$1$-form symmetry and the $\theta$ periodicity with the lattice regularization. 

We note that this study is a non-Abelian extension of the previous study about
$U(1)$ lattice gauge theory~\cite{Abe:2022nfq}, while the actual construction
in this paper is based on a different idea from that
of~Ref.~\cite{Abe:2022nfq}. Still, applying this construction to the $U(1)$
gauge group should reproduce the result of~Ref.~\cite{Abe:2022nfq} due to the
uniqueness of the principal bundle with given data.

\section{Lattice $SU(N)$ gauge theories coupled with $\mathbb{Z}_N$ $2$-form
gauge fields}
\label{sec:ZN_two_form}

Approximating the closed Euclidean spacetime~$M$ as a lattice, we introduce the
$SU(N)$-valued link variables $U_\ell\in SU(N)$ for each link~$\ell$ and the
$\mathbb{Z}_N$-valued plaquette variables $\rme^{\frac{2\pi\im}{N}B_p}$ for
each plaquette~$p$. For the purpose of numerical simulations, one usually takes
$M=T^4$ as the Euclidean spacetime and approximates it as the hypercubic
lattice $\Lambda_L=(\mathbb{Z}/L\mathbb{Z})^4$ since the hypercubic symmetry restricts possible UV divergences and simplifies renormalization procedures. In
this section, let us consider more generic cases for the discussion of general
properties, although we shall also restrict to the hypercubic case later for
the construction of a lattice topological charge. Following the nomenclature
in~Ref.~\cite{Luscher:1981zq}, the $0$-cells are referred to as sites, the
$1$-cells are links, the $2$-cells are plaquettes, the $3$-cells are faces, and
the $4$-cells are just cells. 

When $\ell$ and~$\ell'$ describe the same link with the opposite orientation,
$\ell'=-\ell$, we require that $U_{-\ell}=U_\ell^{-1}$. Similarly,
$B_{-p}=-B_p\bmod N$. Along a plaquette~$p$, we define the $SU(N)$-valued
plaquette variable by the path-ordered product of link variables, 
\begin{equation}
   U_p:=\mathcal{P}\prod_{\ell\in\partial p}U_\ell,
\end{equation}
where $\ell\in\partial p$ refers to the link in the positive relative
orientation for the plaquette~$p$. We require that the $\mathbb{Z}_N$ plaquette
variables satisfy the flatness condition,
\begin{equation}
   \sum_{p\in\partial f}B_p=0\bmod N,
\label{eq:ZN_flatness}
\end{equation}
for each face~$f$. We define the Wilson plaquette action as
\begin{equation}
   S_{\mathrm{W}}[U_\ell,B_p]
   =\sum_p\beta
   \left[\tr\left(\bm{1}-\rme^{-\frac{2\pi\im}{N} B_p}U_p\right)
   +\mathrm{c.c.}\right],
\label{eq:Wilson_action}
\end{equation}
where $\beta$ is the coupling constant for the lattice Yang--Mills theory.

This Wilson action is invariant under the $SU(N)$ gauge transformation: For the
link~$\ell$ connecting two sites $x$ and~$y$, which we denote as
$\ell=\langle x\to y\rangle$, the $SU(N)$ transformation is given by
\begin{equation}
   U_\ell\mapsto g_x^{-1}U_\ell g_y,
\label{eq:SUN_gauge_trans}
\end{equation}
with~$g_{x}\in SU(N)$. Let us denote $n$ as the initial and final point of the
closed loop for~$p$, then $U_p\mapsto g_n^{-1}U_pg_n$ under the $SU(N)$ gauge
transformation so that $\tr(U_p)$ is invariant.

The Wilson action with~$\mathbb{Z}_N$ plaquette fields~$B_p$ is invariant also
under the $\mathbb{Z}_N$ $1$-form gauge transformation, defined by
\begin{align}
   U_\ell&\mapsto\rme^{\frac{2\pi\im}{N}\lambda_\ell}U_\ell,
\notag\\
   B_p&\mapsto B_p +(\diff\lambda)_p\quad\bmod N,
\label{eq:ZN_1form_gauge}
\end{align}
where $\lambda_\ell\in\mathbb{Z}_N$
and~$(\diff\lambda)_p=\sum_{\ell\in\partial p}\lambda_\ell$. We note that the
flatness condition~\eqref{eq:ZN_flatness}, or $(\diff B)_f=0\bmod N$, is
invariant under the $\mathbb{Z}_N$ $1$-form gauge transformations. When $M$ has
no torsion, we can think of $B_p$ as a closed surface on the dual lattice using
the Poincar\'e duality, and the $\mathbb{Z}_N$ $1$-form gauge invariance
requires the invariance under its continuous deformation and the
addition/subtraction of contractible closed surfaces.

In order to introduce the notion of topological sectors for the lattice gauge
fields, the admissibility condition~\cite{Luscher:1981zq} plays the key role.
We say that the $SU(N)$ gauge field~$\{U_\ell\}$ is admissible, if and only if,
for a given $0<\varepsilon\ll2$, all the plaquettes $U_p$ satisfy
\begin{equation}
   \left\|\bm{1}-\rme^{-\frac{2\pi\im}{N}B_p}U_p\right\|<\varepsilon.
\label{eq:admissibilty}
\end{equation}
Here, $\|\cdot\|$ refers the matrix norm. After diagonalization
$\rme^{-\frac{2\pi\im}{N}B_p}U_p=V\rme^{\diag(\im\theta_1,\dotsc,\im\theta_N)}V^{-1}$
with~$\theta_1+\dotsb+\theta_N=0$, we find that
$\|\bm{1}-\rme^{-\frac{2\pi\im}{N}B_p}U_p\|
=\max_k\left|2\sin(\theta_k/2)\right|<\varepsilon$. We denote the set of
admissible gauge fields as
\begin{equation}
   \mathfrak{A}_{\varepsilon}[B_p]
   =\left\{
   \left\{U_\ell\right\}\,
   \Bigm|
   \text{$\left\|\bm{1}-\rme^{-\frac{2\pi\im}{N}B_p}U_p\right\|<\varepsilon$ for
   all $p$}
   \right\},
\end{equation} 
and $\mathfrak{A}_{\varepsilon}[B_p]$ is $SU(N)$ gauge invariant and
$\mathbb{Z}_N$ $1$-form gauge covariant. The functional integral is performed
over~$\mathfrak{A}_{\varepsilon}[B_p]$ so that the partition function is given
as\footnote{Here, we introduce the sharp cutoff for the integration domain. One
can instead introduce the bump function to make everything smooth and it would
ensure better analytic properties of the partition function.}
\begin{equation}
   \calZ[B_p]
   =\int_{\mathfrak{A}_\varepsilon[B_p]}
   \Diff U_\ell\,
   \exp\left(-S_{\mathrm{W}}[U_\ell, B_p]+\dotsb\right),
\end{equation} 
where $\Diff U_\ell$ is the product of the $SU(N)$ Haar measure for all the
link variables. We note that $\varepsilon$ shall be determined independently
of the lattice size, coupling constant~$\beta$, and the plaquette gauge
fields~$B_p$. It will be discussed in~Appendix~\ref{sec:bound_epsilon} and we
find~$\ve\lesssim0.074$ is sufficiently small.

\section{Transition functions from lattice gauge fields}
\label{sec:transition_func}

In this section, we construct the transition function for admissible lattice
gauge fields coupled to the $\mathbb{Z}_N$ $2$-form gauge fields~$B_p$
extending the seminal work by~L\"uscher~\cite{Luscher:1981zq}. We work on the
$4$-torus $M=T^4$ and its hypercubic lattice
discretization~$\Lambda_L=(\mathbb{Z}/L\mathbb{Z})^4$, i.e., each direction is
cut into $L$ pieces. As illustrated in~Fig.~\ref{fig:def_variables}, we define
some variables on the hypercubic structure as follows: The unit cell is
denoted as
\begin{equation}
   c(n)
   =\left\{x\in T^4\bigm|
   \text{$0\le x_\mu-n_\mu\le1$ for $\mu=1$, \dots, $4$}\right\}
\end{equation} 
for~$n\in\Lambda_L$, and we would like to define the transition function on
each face
\begin{equation}
   f(n,\mu)=c(n)\cap c(n-\Hat{\mu}),
\end{equation} 
where $\Hat{\mu}$ is the unit vector along the $\mu$th direction.

For the link~$\ell=\langle n\to n+\Hat{\mu}\rangle$, we write the link variable
as~$U(n,\mu)=U_\ell$. For the plaquette including $n$, $n+\Hat{\mu}$,
$n+\Hat{\nu}$, and $n+\Hat{\mu}+\Hat{\nu}$, we have the $SU(N)$ plaquette
variable
\begin{equation}
   U_{\mu\nu}(n):=U(n,\mu)U(n+\Hat{\mu},\nu)
   U(n+\Hat{\nu},\mu)^{-1}U(n,\nu)^{-1},
\end{equation}
and $\mathbb{Z}_N$ $2$-form gauge field~$B_{\mu\nu}(n)=B_p$. The $1$-form gauge
invariant plaquette is then given by
\begin{equation}
   \Tilde{U}_{\mu\nu}(n):=\rme^{-\frac{2\pi\im}{N}B_{\mu\nu}(n)}U_{\mu\nu}(n),
\end{equation}
and the admissibility condition~\eqref{eq:admissibilty} requires
$\|\bm{1}-\Tilde{U}_{\mu\nu}(n)\|<\varepsilon$ for all the plaquettes.

\subsection{Complete axial gauge and transition functions at the corner}
\label{sec:complete_axial}

\begin{figure}
    \centering
    \begin{tikzpicture}[scale=0.5]
        \draw (0,0,0)--(0,4,0)--(4,4,0)--(4,0,0)--(0,0,0);
        \draw (4,0,0)--(4,0,-4)--(4,4,-4)--(4,4,0);
        \draw (4,4,-4)--(0,4,-4)--(0,4,0);
        \draw[dashed](0,0,0)--(0,0,-4)--(0,4,-4);
        \draw[dashed](0,0,-4)--(4,0,-4);
        \node at (0,0,0) [below]{$s_0$};
        \node at (4,0,0)[below]{$s_1$};
        \node at (4,0,-4)[below right]{$s_6$};
        \node at (0,0,-4)[below right]{$s_2$};
        \node at (4,4,0) [right] {$s_5$};
        \node at (4,4,-4)[right]{$s_4$};
        \node at (0,4,-4)[above]{$s_7$};
        \node at (0,4,0)[left]{$s_3$};
        \draw[->,>=stealth] (-3,0,0) -- (-1.5,0,0) node [right]{$x_\alpha$};
        \draw[->,>=stealth] (-3,0,0) -- (-3,0,-2) node [above]{$x_\beta$};
        \draw[->,>=stealth] (-3,0,0) -- (-3,2,0) node [above]{$x_\gamma$};
    \end{tikzpicture}
\caption{We introduce the coordinate for the face~$f(n,\mu)$
as~$x=n+y_\alpha\Hat{\alpha}+y_\beta\Hat{\beta}+y_\gamma\Hat{\gamma}$
with~$0\le y_{\alpha,\beta,\gamma}\le1$ and~$\alpha<\beta<\gamma$. We label
the corners of~$f(n,\mu)$ (i.e., $y_{\alpha,\beta,\gamma}\in\{0,1\}$) as $s_i$
($i=0$, $1$, \dots, $7$) as shown in this figure.}
\label{fig:face}
\end{figure}
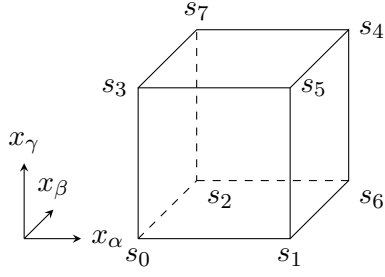

We first define the complete axial gauge for each cell~$c(n)$ and introduce the
transition function at the corner of the cells as the connection formula of
the corresponding link variables. Let $x\in c(n)$ be a corner, i.e.,
$x=n+\sum_\mu z_\mu\Hat{\mu}$ with~$z_\mu\in\{0,1\}$, and we define the standard
parallel transporter from~$n$ to~$x$ as\footnote{We adopt a somewhat different
definition of the standard transporter from L\"uscher's one
in~Ref.~\cite{Luscher:1981zq}. Most part of the discussion goes through in the
same manner for both conventions, except for the explicit formulas
of~$\Tilde{u}_{xy}^m$ and~$\Tilde{v}_{n,\mu}(x)$.}
\begin{equation}
   w^n(x)
   :=U(n,4)^{z_4}U(n+z_4\Hat{4},3)^{z_3}
   U(n+z_4\Hat{4}+z_3\Hat{3},2)^{z_2}
   U(n+z_4\Hat{4}+z_3\Hat{3}+z_2\Hat{2},1)^{z_1}.
\label{eq:standard_transporter}
\end{equation}
We then set, for two adjacent corners $x$ and~$y$ of~$c(n)$,
\begin{align}
   u_{xy}^n&:=w^n(x)U(x,\mu)w^n(y)^{-1},&&\text{if $y=x+\Hat{\mu}$},
\notag\\
   u_{xy}^n&:=\left(u_{yx}^n\right)^{-1},&&\text{if $y=x-\Hat{\mu}$}.
\label{eq:complete_axial}
\end{align}
This can be thought of the link variable in the complete axial gauge of~$c(n)$
for~$\ell=\langle x\to y\rangle$, since all the link variables along the
standard parallel transporter are set to be~$\bm{1}$. When the link
$\ell=\langle x\to y\rangle$ is shared by two cells $c(n)$
and~$c(n-\Hat{\mu})$, the link variables $u_{xy}^n$ and~$u_{xy}^{n-\Hat{\mu}}$
take different values due to the different choice of the gauge between these
cells, and we introduce the transition functions at the corner $v_{n,\mu}(x)$
to make the connection between them:
\begin{equation}
   u^{n-\Hat{\mu}}_{xy}=v_{n,\mu}(x)u^{n}_{xy}v_{n,\mu}(y)^{-1}.
\end{equation}
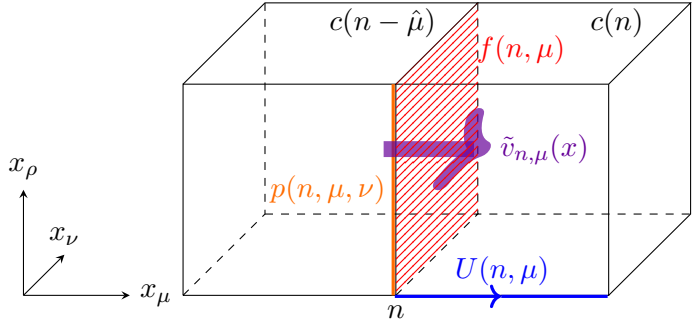
\begin{figure}
    \centering
    \begin{tikzpicture}[scale=0.7]
        \draw (0,0,0)--(0,4,0)--(4,4,0)--(4,0,0)--(0,0,0);
        \draw (4,4,0)--(4,4,-4)--(0,4,-4)--(0,4,0);
        \draw (4,0,0)--(8,0,0)--(8,4,0);
        \draw (4,4,0)--(8,4,0)--(8,4,-4)--(4,4,-4);
        \draw (8,0,0)--(8,0,-4)--(8,4,-4);
        \draw[dashed](0,0,0)--(0,0,-4)--(0,4,-4);
        \draw[dashed](0,0,-4)--(4,0,-4)--(4,0,0);
        \draw[dashed](4,4,-4)--(4,0,-4)--(8,0,-4);
        \draw [red!15!orange, very thick] (3.95,0,0)--(3.95,4,0);
        \draw [->,blue, very thick] (4,0,0.05)--(6,0,0.05);
        \draw [blue, very thick] (6,0,0.05)--(8,0,0.05);
        \draw [red!40!blue, opacity=0.7, arrows = {->[slant=0.5]}, line width = 6] (3,2,-2) -- (5,2,-2);
        \fill[pattern=north east lines, pattern color = red](4,0,0)--(4,0,-4)--(4,4,-4)--(4,4,0);
        \node [red] at (5.2,3.5,-3) {$f(n,\mu)$};
        \node at (4,0,0) [below] {$n$};
        \node [red!15!orange] at (4,2,0) [left] {$p(n,\mu,\nu)$};
        \node [blue] at (6,0,0) [above] {$U(n,\mu)$};
        \node [red!40!blue] at (5,2,-2) [right] {$\Tilde{v}_{n,\mu}(x)$};
        \node at  (2.6,4,-3) {$c(n-\Hat{\mu})$};
        \node at (7,4,-3) {$c(n)$};
        \draw[->,>=stealth] (-3,0,0) -- (-1,0,0) node [right]{$x_\mu$};
        \draw[->,>=stealth] (-3,0,0) -- (-3,0,-2) node [above]{$x_\nu$};
        \draw[->,>=stealth] (-3,0,0) -- (-3,2,0) node [above]{$x_\rho$};
    \end{tikzpicture}
\caption{Illustration for the definition of variables. The transition
functions~$\Tilde{v}_{n,\mu}(x)$ ($1$-form covariant counterpart
of~$v_{n,\mu}(x)$) are defined on the face~$f(n,\mu)=c(n)\cap c(n-\Hat{\mu})$,
and they are subject to the cocycle condition
on~$p(n,\mu,\nu)=%
c(n)\cap c(n-\Hat{\mu})\cap c(n-\Hat{\mu})\cap c(n-\Hat{\mu}-\Hat{\nu})$. At
this moment, $\Tilde{v}_{n,\mu}(x)$ is defined only at the corner
of~$f(n,\mu)$, and we later interpolate it.}
\label{fig:def_variables}
\end{figure}
The explicit formula for $v_{n,\mu}(x)$ is given by
\begin{equation}
   v_{n,\mu}(x)=w^{n-\Hat{\mu}}(x)w^n(x)^{-1}, 
\end{equation}
when $x$ is at the corner of~$f(n,\mu)=c(n)\cap c(n-\Hat{\mu})$. For later
purpose, we denote the corner of~$f(n,\mu)$ as~$s_i$ ($i=0$, $1$, \dots, $7$)
as illustrated in~Fig.~\ref{fig:face}, and we introduce the coordinate
on~$f(n,\mu)$ 
as~$x=n+y_\alpha\Hat{\alpha}+y_\beta\Hat{\beta}+y_\gamma\Hat{\gamma}$
with~$\alpha$, $\beta$, $\gamma\in\{1,2,3,4\}\setminus\{\mu\}$
and~$\alpha<\beta<\gamma$.

\begin{figure}
    \centering
    \begin{tikzpicture}[scale=0.95]
        \foreach \x in {0,1}{
            \foreach \y in {0,1}{
                \draw (0-4*\x+8*\y,0-6*\x,0)--(0-4*\x+8*\y,2-6*\x,0)--(2-4*\x+8*\y,2-6*\x,0)--(2-4*\x+8*\y,0-6*\x,0)--(0-4*\x+8*\y,0-6*\x,0);
                \draw (2-4*\x+8*\y,0-6*\x,0)--(2-4*\x+8*\y,0-6*\x,-2)--(2-4*\x+8*\y,2-6*\x,-2)--(2-4*\x+8*\y,2-6*\x,0);
                \draw (2-4*\x+8*\y,2-6*\x,-2)--(0-4*\x+8*\y,2-6*\x,-2)--(0-4*\x+8*\y,2-6*\x,0);
                \draw[dashed](0-4*\x+8*\y,0-6*\x,0)--(0-4*\x+8*\y,0-6*\x,-2)--(0-4*\x+8*\y,2-6*\x,-2);
                \draw[dashed](0-4*\x+8*\y,0-6*\x,-2)--(2-4*\x+8*\y,0-6*\x,-2);
            };
        };
        \begin{scope}[decoration={markings,mark=at position 0.5 with {\arrow{>}}}] 
            \draw [very thick,postaction={decorate},orange] (-4,-6,0)--(-4,-4,0);
            \draw [very thick,postaction={decorate},orange] (-4,-4,0) .. controls (-3,-3,0) and (-1,3,0) .. (0,2,0);
            \draw [very thick,postaction={decorate},red]  (0,0,0) .. controls (-1,1,0) and (-3,-5,0)..(-4,-6,0);
            \draw [very thick,postaction={decorate},orange] (-4,-6,0)--(-4,-4,0);
            \draw [very thick,postaction={decorate},orange] (0,2,0)--(0,2,-2);
            \draw [very thick,postaction={decorate},blue] (0,2,-2)--(0,0,-2);
            \draw [very thick,postaction={decorate},red] (0,0,-2)--(0,0,0);
        \end{scope}
        \begin{scope}[decoration={markings,mark=at position 0.5 with {\arrow{>}}}] 
            \draw [very thick,postaction={decorate}] (4,-6,0)--(4,-4,0);
            \draw [very thick,postaction={decorate}] (4,-4,0) .. controls (5,-3,0) and (7,3,0) .. (8,2,0);
            \draw [very thick,postaction={decorate}]  (8,0,0) .. controls (7,1,0) and (5,-5,0)..(4,-6,0);
            \draw [very thick,postaction={decorate}] (4,-6,0)--(4,-4,0);
            \draw [very thick,postaction={decorate}] (8,2,0)--(8,2,-2);
            \draw [very thick,postaction={decorate}] (8,2,-2)--(8,0,-2);
            \draw [very thick,postaction={decorate}] (8,0,-2)--(8,0,0);
        \end{scope}
        \path [fill, pattern=north east lines, pattern color = orange] plot (4,-4,0) .. controls (5,-3,0) and (7,3,0) .. (8,2,0)--plot (8,0,0)-- plot (8,0,0) .. controls (7,1,0)  and (5,-5,0)  .. (4,-6,0)--plot(4,-6,0);
        \fill[pattern=north east lines, pattern color = red](8,0,0)--(8,0,-2)--(8,2,-2)--(8,2,0);
        \node [red] at (-0.5,-1.5,0){$w^{n-\Hat{3}}(s_2)^{-1}$};
        \node [red!15!orange] at (-3.5,-1,0){$w^{n-\Hat{3}}(s_7)$};
        \node [red] at (8.8,0.8,-2){$B_{24}(n)$};
        \node [red!15!orange] at (7.5,-1.5,0){$B_{34}(n-\Hat{3})$};
        \node [blue] at (0,0.5,-2) [right]{$U(s_7,-4)$};
        \node at (-4,-6,0) [below]{$n-\Hat{3}$};
        \node at (0,0,0) [below]{$s_0=n$};
        \node at (2,0,0)[below]{$s_1$};
        \node at (2,0,-2)[below right]{$s_6$};
        \node at (0,0,-2)[below right]{$s_2$};
        \node at (2,2,0) [right] {$s_5$};
        \node at (2,2,-2)[right]{$s_4$};
        \node at (0,2,-2)[above]{$s_7$};
        \node at (0,1.8,0)[left]{$s_3$};
        \draw[->,>=stealth] (-4,0,0) -- (-3,0,0) node [right]{$x_1$};
        \draw[->,>=stealth] (-4,0,0) -- (-4,0,-1) node [above]{$x_2$};
        \draw[->,>=stealth] (-4,0,0) -- (-4,1,0) node [above]{$x_4$};
        \node at (-0.5,-7,0) {Construction of $u_{s_7s_2}^{n-\Hat{3}}$};
        \node at (7.5,-7,0) {Construction of $\Tilde{u}_{s_7s_2}^{n-\Hat{3}}$};
    \end{tikzpicture}
\caption{
Illustration for the construction of~$u_{xy}^m$ (left) and its $1$-form
invariant counterpart $\Tilde{u}_{xy}^m$ (right). We take an example
with~$m=n-\Hat{3}$, $x=s_7=n+\Hat{2}+\Hat{4}$, and~$y=s_2=n+\Hat{2}$, so we
draw two cubes for faces~$f(n,3)$ and~$f(n-\Hat{3},3)$.
\textbf{(Left)}~Each building block
of~$u_{s_7s_2}^{n-\Hat{3}}=w^{n-\Hat{3}}(s_7)U(s_7,-4)w^{n-\Hat{3}}(s_2)^{-1}$
is shown in different colors, in which the links in the $\Hat{3}$ direction are
drawn with curved lines.
\textbf{(Right)}~As $u_{xy}^m$ forms a closed loop starting and ending at the
site~$m$, we attach the $\mathbb{Z}_N$ plaquette fields
$\rme^{\frac{2\pi\im}{N}B_{\mu\nu}(n)}$ when the loop surrounds the 
corresponding plaquettes.}
\label{fig:construction_tildeU}
\end{figure}
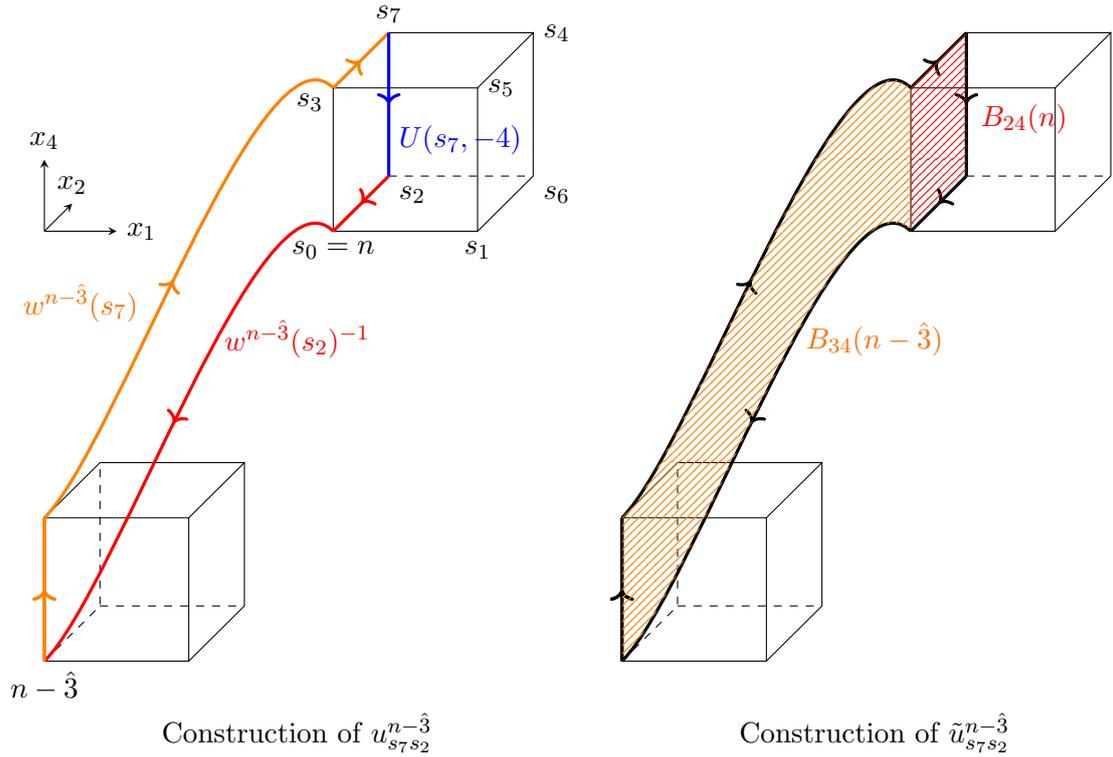

The link in the complete axial gauge $u_{xy}^m$ is the product of link
variables along a closed loop that starts and ends at the site~$m$. Therefore,
we can make them invariant under the $1$-form gauge transformation by
multiplying appropriate factors of~$\rme^{-\frac{2\pi\im}{N}B_{\mu\nu}(n)}$.
Due to the flatness condition~\eqref{eq:ZN_flatness}, there is a unique local
way to achieve the $\mathbb{Z}_N$ $1$-form gauge invariance, and such $1$-form
invariant $u_{xy}^m$ are denoted with the tilde, $\Tilde{u}_{xy}^m$. This
construction is illustrated in~Fig.~\ref{fig:construction_tildeU} by taking an
example of~$\Tilde{u}_{s_7s_2}^{m=n-\Hat{3}}$, where $s_7$ and~$s_2$ denote
corners of~$f(n,3)$, and it shows that
\begin{equation}
   \Tilde{u}_{s_7s_2}^{n-\Hat{3}}
   =\rme^{\frac{2\pi\im}{N}B_{34}(n-\Hat{3})}\rme^{\frac{2\pi\im}{N}B_{24}(n)}
   u_{s_7s_2}^{n-\Hat{3}}. 
\end{equation}
The list of the concrete expressions for~$\Tilde{u}_{xy}^m$ is given
in~Appendix~\ref{sec:formula_axial_gauge}.

We further want to define $\Tilde{v}_{n,\mu}(x)$ for a corner~$x$ of~$f(n,\mu)$
such that
\begin{equation}
   \Tilde{u}_{xy}^{n-\Hat{\mu}}
   =\Tilde{v}_{n,\mu}(x)\Tilde{u}_{xy}^n\Tilde{v}_{n,\mu}(y)^{-1},
\label{eq:connection_tilded}
\end{equation}
where $x$ and~$y$ are the corners of~$f(n,\mu)$. Also, we impose that
$\Tilde{v}_{n,\mu}(x)$ for the corner~$x$ behaves covariantly under the
$1$-form gauge transformation,
\begin{equation}
   \Tilde{v}_{n,\mu}(x)
   \mapsto\rme^{\frac{2\pi\im}{N}\lambda_\mu(n-\Hat{\mu})}\Tilde{v}_{n,\mu}(x).
\label{eq:(3.11)}
\end{equation}
This is again achieved by multiplication of appropriate factors
of~$\rme^{\frac{2\pi\im}{N}B_{\mu\nu}(n)}$, and the concrete expressions are
given in~Appendix~\ref{sec:transition_1formcov}. We can then confirm
Eq.~\eqref{eq:connection_tilded} by using the flatness condition,
$(\diff B)_f=0\bmod N$. From these, we observe the cocycle condition at the
corner~$n$ of~$p(n,\mu,\nu)=%
c(n)\cap c(n-\Hat{\mu})\cap c(n-\Hat{\nu})\cap c(n-\Hat{\mu}-\Hat{\nu})$, 
\begin{align}
   \Tilde{v}_{n-\Hat{\nu},\mu}(n)\Tilde{v}_{n,\nu}(n)
   \Tilde{v}_{n,\mu}(n)^{-1}\Tilde{v}_{n-\Hat{\mu},\nu}(n)^{-1}
   =\rme^{\frac{2\pi \im}{N}B_{\mu\nu}(n-\Hat{\mu}-\Hat{\nu})}\bm{1}.
\label{eq:cocycle_corner}
\end{align}
We can check it using the relation
$v_{n-\Hat{\nu},\mu}(n)v_{n,\nu}(n)=v_{n-\Hat{\mu},\nu}(n)v_{n,\mu}(n)$ and the
formulas given in~Appendix~\ref{sec:transition_1formcov}. This is the desired
cocycle condition in the $SU(N)/\mathbb{Z}_N$ gauge theory and actually is
expected from the transformation property under the $1$-form gauge
transformation alone: From the cocycle condition for transition
functions without tildes, the right-hand side must be an element of~$\mathbb{Z}_N$. This
$\mathbb{Z}_N$ element should transform identically as the left-hand side does under
the $\mathbb{Z}_N$ $1$-form gauge transformation~\eqref{eq:(3.11)}. Then, the
unique possibility is
$\rme^{\frac{2\pi \im}{N}B_{\mu\nu}(n-\Hat{\mu}-\Hat{\nu})}\bm{1}$.

\subsection{Continuum interpolation of transition functions and cocycle
conditions}

So far, we have defined the transition functions~$\Tilde{v}_{n,\mu}(x)$ only at
the corners of~$f(n,\mu)$. In this section, we define the smooth transition
functions
\begin{equation}
   \Tilde{v}_{n,\mu}:f(n,\mu)\to SU(N),
\end{equation}
by introducing the interpolation functions~$\Tilde{S}^m_{n,\mu}(x)$ so that 
\begin{equation}
   \Tilde{v}_{n,\mu}(x)\equiv
   \Tilde{S}_{n,\mu}^{n-\Hat{\mu}}(x)^{-1}\Tilde{v}_{n,\mu}(n)
   \Tilde{S}_{n,\mu}^n(x). 
\end{equation}
In order for this to be an interpolation, the iterative application
of~Eq.~\eqref{eq:connection_tilded} implies that
$\Tilde{S}^{m}_{n,\mu}:f(n,\mu)\to SU(N)$ should satisfy
\begin{equation}
   \Tilde{S}_{n,\mu}^m(x)
   =\begin{cases}
   1,
   &\text{for $x=s_0$, i.e., $(y_\alpha,y_\beta,y_\gamma)=(0,0,0)$},\\
   \Tilde{u}_{s_0s_1}^m,
   &\text{for $x=s_1$, i.e., $(y_\alpha,y_\beta,y_\gamma)=(1,0,0)$},\\
   \Tilde{u}_{s_0s_2}^m,
   &\text{for $x=s_2$, i.e., $(y_\alpha,y_\beta,y_\gamma)=(0,1,0)$},\\
   \Tilde{u}_{s_0s_3}^m,
   &\text{for $x=s_3$, i.e., $(y_\alpha,y_\beta,y_\gamma)=(0,0,1)$},\\
   \Tilde{u}_{s_0s_3}^m\Tilde{u}_{s_3s_7}^m\Tilde{u}_{s_7s_4}^m,
   &\text{for $x=s_4$, i.e., $(y_\alpha,y_\beta,y_\gamma)=(1,1,1)$},\\
   \Tilde{u}_{s_0s_3}^m\Tilde{u}_{s_3s_5}^m,
   &\text{for $x=s_5$, i.e., $(y_\alpha,y_\beta,y_\gamma)=(1,0,1)$},\\
   \Tilde{u}_{s_0s_2}^m\Tilde{u}_{s_2s_6}^m,
   &\text{for $x=s_6$, i.e., $(y_\alpha,y_\beta,y_\gamma)=(1,1,0)$},\\
   \Tilde{u}_{s_0s_3}^m\Tilde{u}_{s_3s_7}^m,
   &\text{for $x=s_7$, i.e., $(y_\alpha,y_\beta,y_\gamma)=(0,1,1)$}.\\
   \end{cases}
\label{eq:S_corner}
\end{equation}
Another important requirement is that this interpolated function should satisfy
the cocycle condition, that is, for~$x\in p(n,\mu,\nu)$,
\begin{equation}
   \Tilde{v}_{n-\Hat{\nu},\mu}(x)\Tilde{v}_{n,\nu}(x)
   \Tilde{v}_{n,\mu}(x)^{-1}\Tilde{v}_{n-\Hat{\mu},\nu}(x)^{-1}
   =\rme^{\frac{2\pi \im}{N}B_{\mu\nu}(n-\Hat{\mu}-\Hat{\nu})}\bm{1},
\label{eq:cocycle}
\end{equation}
which gives a nontrivial condition for~$\Tilde{S}_{n,\mu}^m(x)$.

Such an interpolation $\Tilde{S}_{n,\mu}^m(x)$ has been constructed explicitly
by~L\"uscher in~Eqs.~(22a)--(22f) of~Ref.~\cite{Luscher:1981zq} for the case
of~$SU(N)$. We extend the formula to the case of~$SU(N)/\mathbb{Z}_N$ by
imposing the manifest $1$-form gauge invariance. The explicit form
of~$\Tilde{S}_{n,\mu}^m(x)$ is written as follows:
\begin{align}
   \Tilde{f}_{n,\mu}^m(x_\gamma)
   &=(\Tilde{u}_{s_3s_0}^m)^{y_\gamma}
   (\Tilde{u}_{s_0s_3}^m\Tilde{u}_{s_3s_7}^m\Tilde{u}_{s_7s_2}^m\Tilde{u}_{s_2s_0}^m
   )^{y_\gamma}
   \Tilde{u}_{s_0s_2}^m
   (\Tilde{u}_{s_2s_7}^m)^{y_\gamma},
\notag\\
   \Tilde{g}_{n,\mu}^m(x_\gamma)
   &=(\Tilde{u}_{s_5s_1}^m)^{y_\gamma}
   (\Tilde{u}_{s_1s_5}^m\Tilde{u}_{s_5s_4}^m\Tilde{u}_{s_4s_6}^m\Tilde{u}_{s_6s_1}^m
   )^{y_\gamma}
   \Tilde{u}_{s_1s_6}^m
   (\Tilde{u}_{s_6s_4}^m)^{y_\gamma},
\notag\\
   \Tilde{h}_{n,\mu}^m(x_\gamma)
   &=(\Tilde{u}_{s_3s_0}^m)^{y_\gamma}
   (\Tilde{u}_{s_0s_3}^m\Tilde{u}_{s_3s_5}^m\Tilde{u}_{s_5s_1}^m\Tilde{u}_{s_1s_0}^m
   )^{y_\gamma}
   \Tilde{u}_{s_0s_1}^m
   (\Tilde{u}_{s_1s_5}^m)^{y_\gamma},
\notag\\
   \Tilde{k}_{n,\mu}^m(x_\gamma)
   &=(\Tilde{u}_{s_7s_2}^m)^{y_\gamma}
   (\Tilde{u}_{s_2s_7}^m\Tilde{u}_{s_7s_4}^m\Tilde{u}_{s_4s_6}^m\Tilde{u}_{s_6s_2}^m
   )^{y_\gamma}
   \Tilde{u}_{s_2s_6}^m
   (\Tilde{u}_{s_6s_4}^m)^{y_\gamma},
\notag\\
   \Tilde{l}_{n,\mu}^m(x_\beta,x_\gamma)
   &=\left[\Tilde{f}_{n,\mu}^m(x_\gamma)^{-1}\right]^{y_\beta}
   \left[
   \Tilde{f}_{n,\mu}^m(x_\gamma)\Tilde{k}_{n,\mu}^m(x_\gamma)
   \Tilde{g}_{n,\mu}^m(x_\gamma)^{-1}\Tilde{h}_{n,\mu}^m(x_\gamma)^{-1}
   \right]^{y_\beta}
\notag\\
   &\qquad{}
   \cdot
   \Tilde{h}_{n,\mu}^m(x_\gamma)
   \left[\Tilde{g}_{n,\mu}^m(x_\gamma)\right]^{y_\beta},
\notag\\
   \Tilde{S}_{n,\mu}^m(x_\alpha,x_\beta,x_\gamma)
   &=(\Tilde{u}_{s_0s_3}^m)^{y_\gamma}
   \left[\Tilde{f}_{n,\mu}^m(x_\gamma)\right]^{y_\beta}
   \left[\Tilde{l}_{n,\mu}^m(x_\beta,x_\gamma)\right]^{y_\alpha}.
\label{eq:interpolater}
\end{align}
In these formulas, we need to take the fractional power of plaquette variables
as defined in~Appendix~\ref{sec:fractional_power}, and their well-definedness is
ensured by imposing the admissibility $\{U_\ell\}\in\mathfrak{A}_{\ve}[B_p]$ for
sufficiently small~$\ve\lesssim0.074$. We can confirm that, when $x$~is one of
the corners of~$f(n,\mu)$, this definition reduces to~Eq.~\eqref{eq:S_corner}
as requested. 
Moreover, $\tilde{v}_{n,\mu}(x)$ obeys the $1$-form transformation proeprety \eqref{eq:(3.11)} since $\tilde{S}^{m}_{n,\mu}$ is $1$-form gauge invariant. 

Let us now check if the above $\Tilde{v}_{n,\mu}(x)$ satisfies the cocycle
condition~\eqref{eq:cocycle}. It turns out that L\"uscher's proof for the Lemma
in~Ref.~\cite{Luscher:1981zq} works in the same manner by making its every step manifestly $1$-form
gauge invariant. We first note that
$\Tilde{S}_{n,\mu}^m(x)=\Tilde{S}_{n,\lambda}^m(x)$ when $x\in p(n,\mu,\lambda)$
by construction, and we denote it as
\begin{equation}
   \Tilde{P}_{n,\mu\lambda}^m(x)
   :=\Tilde{S}_{n,\mu}^m(x)=\Tilde{S}_{n,\lambda}^m(x)\qquad
   \text{for $x\in p(n,\mu,\lambda)$}.
\end{equation}
Following Ref.~\cite{Luscher:1981zq}, it is also convenient to introduce
\begin{equation}
   \Tilde{R}_{n,\mu;\lambda}^m(x)
   =\Tilde{S}_{n,\mu}^m(x)\Tilde{P}_{n+\Hat{\lambda},\mu\lambda}^m(x)^{-1}
   \qquad
   \text{for $x\in p(n+\Hat{\lambda},\mu,\lambda)$},
\end{equation}
and its explicit form is given by
\begin{align}
   \Tilde{R}_{n,\mu;\alpha}^m(x_\beta,x_\gamma)
   &=
   \bigl[
   (\Tilde{u}_{s_0s_3}^m\Tilde{u}_{s_3s_7}^m\Tilde{u}_{s_7s_2}^m\Tilde{u}_{s_2s_0}^m
   )^{y_\gamma}
   \Tilde{u}_{s_0s_2}^m
\notag\\
   &\qquad{}
   \cdot
   (\Tilde{u}_{s_2s_7}^m\Tilde{u}_{s_7s_4}^m\Tilde{u}_{s_4s_6}^m\Tilde{u}_{s_6s_2}^m
   )^{y_\gamma}
   \Tilde{u}_{s_2s_6}^m\Tilde{u}_{s_6s_1}^m
   (\Tilde{u}_{s_1s_6}^m\Tilde{u}_{s_6s_4}^m\Tilde{u}_{s_4s_5}^m\Tilde{u}_{s_5s_1}^m
   )^{y_\gamma}
\notag\\
   &\qquad{}
   \cdot\Tilde{u}_{s_1s_0}^m
   (\Tilde{u}_{s_0s_1}^m\Tilde{u}_{s_1s_5}^m\Tilde{u}_{s_5s_3}^m\Tilde{u}_{s_3s_0}^m
   )^{y_\gamma}
   \bigr]^{y_\beta}
   (\Tilde{u}_{s_0s_3}^m\Tilde{u}_{s_3s_5}^m\Tilde{u}_{s_5s_1}^m\Tilde{u}_{s_1s_0}^m
   )^{y_\gamma}
   \Tilde{u}_{s_0s_1}^m,
\notag\\
   \Tilde{R}_{n,\mu;\beta}^m(x_\alpha,x_\gamma)
   &=
   (\Tilde{u}_{s_0s_3}^m\Tilde{u}_{s_3s_7}^m\Tilde{u}_{s_7s_2}^m\Tilde{u}_{s_2s_0}^m
   )^{y_\gamma}
   \Tilde{u}_{s_0s_2}^m,
\notag\\
   \Tilde{R}_{n,\mu;\gamma}^m(x_\alpha,x_\beta)
   &=\Tilde{u}_{s_0s_3}^m.
\end{align}
Using the crucial relation~\eqref{eq:connection_tilded} combined with the
formula~\eqref{eq:conjugate}, we obtain
\begin{equation}
   \Tilde{R}_{n,\mu;\lambda}^{n-\Hat{\mu}}(x)
   =\Tilde{v}_{n,\mu}(n)
   \Tilde{R}_{n,\mu;\lambda}^n(x)
   \Tilde{v}_{n,\mu}(n+\Hat{\lambda})^{-1}.
\end{equation}
These show that, for~$x\in p(n,\mu,\lambda)$,
\begin{equation}
   \Tilde{v}_{n,\mu}(x)
   =\Tilde{P}_{n,\mu\lambda}^{n-\Hat{\mu}}(x)^{-1}
   \Tilde{v}_{n,\mu}(n)\Tilde{P}_{n,\mu\lambda}^n(x),
\end{equation}
and, for $x\in p(n+\Hat{\lambda},\mu,\lambda)$,
\begin{align}
   \Tilde{v}_{n,\mu}(x)
   &=\Tilde{P}_{n+\Hat{\lambda},\mu\lambda}^{n-\Hat{\mu}}(x)^{-1}
   \Tilde{R}_{n,\mu;\lambda}^{n-\Hat{\mu}}(x)^{-1}
   \Tilde{v}_{n,\mu}(n)
   \Tilde{R}_{n,\mu;\lambda}^n(x)
   \Tilde{P}_{n+\Hat{\lambda},\mu\lambda}^n(x)
\notag\\
   &=\Tilde{P}_{n+\Hat{\lambda},\mu\lambda}^{n-\Hat{\mu}}(x)^{-1}
   \Tilde{v}_{n,\mu}(n+\Hat{\lambda})
   \Tilde{P}_{n+\Hat{\lambda},\mu\lambda}^n(x).
\end{align}
Using these, for~$x\in p(n,\mu,\nu)$, we find
\begin{align}
   \Tilde{v}_{n,\mu}(x)
   &=\Tilde{P}_{n,\mu\nu}^{n-\Hat{\mu}}(x)^{-1}
   \Tilde{v}_{n,\mu}(n)\Tilde{P}_{n,\mu\nu}^n(x),
\notag\\
   \Tilde{v}_{n,\nu}(x)
   &=\Tilde{P}_{n,\mu\nu}^{n-\Hat{\nu}}(x)^{-1}
   \Tilde{v}_{n,\nu}(n)\Tilde{P}_{n,\mu\nu}^n(x),
\notag\\
   \Tilde{v}_{n-\Hat{\nu},\mu}(x)
   &=\Tilde{P}_{n,\mu\nu}^{n-\Hat{\mu}-\Hat{\nu}}(x)^{-1}
   \Tilde{v}_{n-\Hat{\nu},\mu}(n)\Tilde{P}_{n,\mu\nu}^{n-\Hat{\nu}}(x),
\notag\\
   \Tilde{v}_{n-\Hat{\mu},\nu}(x)
   &=\Tilde{P}_{n,\mu\nu}^{n-\Hat{\mu}-\Hat{\nu}}(x)^{-1}
   \Tilde{v}_{n-\Hat{\mu},\nu}(n)\Tilde{P}_{n,\mu\nu}^{n-\Hat{\mu}}(x),
\end{align}
where we have noted $\Tilde{P}_{n,\nu\mu}^m(x)=\Tilde{P}_{n,\mu\nu}^m(x)$ by
definition. Therefore, the cocycle condition along~$p(n,\mu,\nu)$ yields
\begin{align}
   &\Tilde{v}_{n-\Hat{\nu},\mu}(x)\Tilde{v}_{n,\nu}(x)
   \Tilde{v}_{n,\mu}(x)^{-1}\Tilde{v}_{n-\Hat{\mu},\nu}(x)^{-1}
\notag\\
   &=\Tilde{P}_{n,\mu\nu}^{n-\Hat{\mu}-\Hat{\nu}}(x)^{-1}
   \Tilde{v}_{n-\Hat{\nu},\mu}(n)\Tilde{v}_{n,\nu}(n)
   \Tilde{v}_{n,\mu}(n)^{-1}\Tilde{v}_{n-\Hat{\mu},\nu}(n)^{-1}
   \Tilde{P}_{n,\mu\nu}^{n-\Hat{\mu}-\Hat{\nu}}(x)
\notag\\
   &=\rme^{\frac{2\pi\im}{N}B_{\mu\nu}(n-\Hat{\mu}-\Hat{\nu})}\bm{1},
\end{align}
where we used~Eq.~\eqref{eq:cocycle_corner}. This completes the proof of the
cocycle condition~\eqref{eq:cocycle}.

\section{Fractional topological charge on the lattice and its applications}

We now obtain the $SU(N)/\mathbb{Z}_N$ bundle on~$T^4$ characterized by
transition functions $\Tilde{v}_{n\mu}(x)$ between the cells~$c(n)$
and~$c(n-\Hat{\mu})$. The topological charge, or the second Chern class, is
given by~\cite{Luscher:1981zq,vanBaal:1982ag}
\begin{equation}
   Q_{\mathrm{top}}[U_\ell,B_P]=\sum_{n\in\Lambda_L}q(n),
\end{equation}
where 
\begin{align}
   q(n)
   &=-\frac{1}{24\pi^2}
   \sum_{\mu,\nu,\rho,\sigma}
   \ve_{\mu\nu\rho\sigma}
   \int_{f(n,\mu)}\diff^3x\,
   \tr\left[
   (\Tilde{v}_{n,\mu}^{-1}\p_\nu\Tilde{v}_{n,\mu})
   (\Tilde{v}_{n,\mu}^{-1}\p_\rho\Tilde{v}_{n,\mu})
   (\Tilde{v}_{n,\mu}^{-1}\p_\sigma\Tilde{v}_{n,\mu})
   \right]
\notag\\
   &\qquad{}
   -\frac{1}{8\pi^2}
   \sum_{\mu,\nu,\rho,\sigma}\ve_{\mu\nu\rho\sigma}
   \int_{p(n,\mu,\nu)}\diff^2x\,
   \tr\left[
   (\Tilde{v}_{n,\mu}\p_\rho\Tilde{v}_{n,\mu}^{-1})
   (\Tilde{v}_{n-\Hat{\mu},\nu}^{-1}\p_\sigma\Tilde{v}_{n-\Hat{\mu},\nu})
   \right].
\label{eq:TopologicalChargeDensity}
\end{align}
$Q_{\mathrm{top}}$ introduces the topological classification of admissible
lattice gauge fields~$\mathfrak{A}_{\ve}[B_p]$, and gauge fields in different
sectors cannot be continuously deformed with each other
within~$\mathfrak{A}_{\ve}[B_p]$. We note that this topological charge is
manifestly $SU(N)$ and $\mathbb{Z}_N$ $1$-form gauge invariant. Moreover, since
$\Tilde{v}_{n,\mu}(x)$ is constructed out of the plaquettes~$\Tilde{U}_p$ in
the adjacent cells, the topological charge density is defined in the local way
on the lattice.

\subsection{Fractional shift of the topological charge and mixed 't Hooft
anomaly}

By performing the $1$-form gauge transformations and redefinition of variables,
we find that the $SU(N)/\mathbb{Z}_N$ bundle constructed above is equivalent to
the 't~Hooft twisted boundary condition with the 't~Hooft flux (see, e.g.,
Appendix~A of Ref.~\cite{Tanizaki:2022ngt} for details)
\begin{equation}
   z_{\mu\nu}=\sum_{p\in(T^2)_{\mu\nu}}B_p\qquad\bmod N.
\end{equation}
As shown by van Baal in~Ref.~\cite{vanBaal:1982ag}, we then find that
\begin{equation}
   Q_{\mathrm{top}}
   \in-\frac{1}{N}\frac{\ve_{\mu\nu\rho\sigma}z_{\mu\nu}z_{\rho\sigma}}{8}
   +\mathbb{Z},
\label{eq:vanBaal_fractional}
\end{equation}
and thus the lattice topological charge is shifted by a fractional value in the
unit of~$1/N$. In the above expression, the fractional part may look to be
written using the global data, but it can be written in the local way by using
the cohomological operations:
\begin{equation}
   Q_{\mathrm{top}}
   =-\frac{1}{N}\int_{T^4}\frac{1}{2}P_2(B_p)\qquad\bmod1,
\label{eq:fractionalcharge}
\end{equation}
where the Pontryagin square is defined by
\begin{equation}
    P_2(B_p)=B_p\cup B_p+B_p\cup_1\diff B_p.
\end{equation}
For the definition of (higher-)cup products on the hypercubic lattice, 
see~Ref.~\cite{Chen:2021ppt}. The correction by the $1$-cup product is
introduced so that the $\mathbb{Z}_N$ $1$-form gauge invariance is achieved at
the cochain level. We can then confirm the equivalence between
Eqs.~\eqref{eq:vanBaal_fractional} and~\eqref{eq:fractionalcharge} by
performing the gauge transformation so that $(\diff B)_f=0$ in~$\mathbb{Z}$ and
the Poincar\'e dual of~$B_p$ always has transverse intersections. Then,
$\int\frac{1}{2}P_2(B_p)$ in~Eq.~\eqref{eq:fractionalcharge} counts the number
of intersection points in mod~$N$ and it gives the fractional part
of~Eq.~\eqref{eq:vanBaal_fractional}.

As we have defined the local topological charge, we can introduce it as the
$\theta$~term for the Boltzmann weight of the path integral, and we define the
partition function as
\begin{equation}
   \calZ_{\theta}[B_p]
   =\int_{\mathfrak{A}_{\ve}[B_p]}\Diff U_\ell\,
   \exp\left(
   -S_{\mathrm{W}}[U_\ell,B_p]+\im\theta Q_{\mathrm{top}}[U_\ell,B_p]
   \right).
\end{equation}
In order for the well-definedness of the lattice topological charge, we have to
restrict the domain for the path integral to the admissible gauge fields
$\mathfrak{A}_{\ve}[B_p]$ with some $\ve\lesssim 0.074$.
Using~Eq.~\eqref{eq:fractionalcharge}, we find the following equality for the
lattice Yang--Mills partition function:
\begin{equation}
   \calZ_{\theta+2\pi}[B_p]
   =\exp\left[-\frac{2\pi\im}{N}\int_{T^4}\frac{1}{2}P_2(B_p)\right]
   \calZ_{\theta}[B_p].
\label{eq:lattice_tHooftAnomaly}
\end{equation}
In Ref.~\cite{Gaiotto:2017yup} (see also Refs.~\cite{Tanizaki:2017bam, Komargodski:2017dmc, Kikuchi:2017pcp, Tanizaki:2018xto, Karasik:2019bxn, Cordova:2019jnf}), this is
interpreted as the 't~Hooft anomaly for the $\mathbb{Z}_N$ $1$-form symmetry and
the $\theta$ periodicity of $4$d $SU(N)$ pure Yang--Mills theory, and one can
derive various nontrivial consequences on the Yang--Mills vacua from this
relation. In previous studies, its derivation is based on the classical analysis
of the path-integral measure for smooth gauge field configurations, and it is
assumed that the path integral over quantum gauge fields does not spoil the
relation for classical action. Here, we have given the complete proof of the
anomalous relation~\eqref{eq:lattice_tHooftAnomaly} using the lattice
regularization that maintains the locality and the gauge invariance, and thus
it is now shown to be a rigorous relation for the quantum Yang--Mills theory.

\subsection{Classical continuum limit of the topological charge}

Although we obtained the lattice local expression for the topological charge
density~\eqref{eq:TopologicalChargeDensity}, it is somewhat unrealistic to use
this expression for the practical computations due to the complicated
construction of transition functions~$\Tilde{v}_{n,\mu}(x)$. Therefore, it is
desirable to find a simpler expression for the lattice topological charge. Due
to the quantization of the topological charge, we do not even have to perform
its exact calculation as long as we can control the approximation.

Here, we take the classical continuum approximation for computing the lattice
topological charge and make its connection to the naive continuum formula. Since
$Q_{\mathrm{top}}$ is invariant under the continuous deformation of~$\{U_\ell\}$
as long as the admissibility~\eqref{eq:admissibilty} is satisfied, one can use
the following approximate expression with smeared lattice gauge fields when one ensures (or monitors) the preservation of the admissibility in the smearing
procedures such as the gradient
flow~\cite{Narayanan:2006rf, Luscher:2009eq, Luscher:2010iy, Luscher:2011bx, Bonati:2014tqa}.

To take the classical continuum limit, we formally introduce the lattice
constant~$a$ and take the $a\to0$ limit with fixed torus size~$aL$. When $B_p=0$
everywhere, we can put the assumption that all the link variables are close to
the identity and then the gauge fields are introduced as their phase factor.
When $B_p$ is present, however, this assumption has a self-contradiction as
some link variables necessarily have $O(1)$~deviations from~$\bm{1}$, so we
need more careful treatment about the notion of continuum gauge field. In
contrast, we can safely assume the continuity about the $1$-form
gauge-invariant plaquettes. We introduce the $1$-form gauge invariant field
strength~$\Tilde{F}_{\mu\nu}$ by
\begin{equation}
   \Tilde{U}_{\mu\nu}(n)
   =\exp\left[\im a^2\Tilde{F}_{\mu\nu}(an)\right]
   \simeq\bm{1}+\im a^2\Tilde{F}_{\mu\nu}(an).
\label{eq:FieldStrength}
\end{equation}
For the discussion of continuity, we would like to compare $\Tilde{F}_{\mu\nu}$
at two different lattice points $n$ and~$m$. When $m\in c(n)$, we use the
standard parallel transporter $w^n(m)$ to bring back $\Tilde{F}_{\mu\nu}(am)$ to
the point~$n$, and we say that the gauge field is continuous if
\begin{equation}
   \left\|\Tilde{F}_{\mu\nu}(an)-w^n(m)\Tilde{F}_{\mu\nu}(am)w^n(m)^{-1}
   \right\|\sim\mathcal{O}(a).
\end{equation}
Under this assumption of the continuity, we obtain
\begin{equation}
   \Tilde{u}_{xy}^n
   =1
   -\im\frac{a^2}{2}\sum_{\mu<\nu}(x+y-2n)_\mu(x-y)_\nu\Tilde{F}_{\mu\nu}(an)
   +\mathcal{O}(a^3).
\end{equation}
Then, we substitute this expression to the
definition~\eqref{eq:TopologicalChargeDensity}, and the second term on the
right-hand-side gives the leading contribution. The rest of calculations is
completely identical to the one for $SU(N)$ theories~\cite{Luscher:1981zq}, and
we can find the result by quoting them as
\begin{equation}
   q(n)
   =a^4
   \frac{1}{32\pi^2}\ve_{\mu\nu\rho\sigma}
   \tr\left[\Tilde{F}_{\mu\nu}(an)\Tilde{F}_{\rho\sigma}(an)\right]
   +\mathcal{O}(a^5). 
\end{equation}
The continuum expression for the topological charge is reproduced at the leading
order under the presence of generic $B_p$.

\section{Summary and outlooks}
\label{sec:summary}

We have constructed the topological charge on the lattice for $SU(N)$ gauge
fields coupled with $\mathbb{Z}_N$ $2$-form gauge fields. Introducing the
admissibility condition for the $SU(N)$ link variables with generic flat $B_p$,
we obtain a manifestly local, $SU(N)$ and $\mathbb{Z}_N$ $1$-form
gauge-invariant expression for the topological charge density. As an
application of this result, we provide the rigorous derivation of the mixed
't~Hooft anomaly between the $\mathbb{Z}_N$ $1$-form symmetry and the $\theta$
periodicity. We also give the classical continuum limit of the topological
charge density, as the exact expression is so complicated for the practical use.
We introduce the gauge-invariant criterion for the validity of continuum
approximation, so one can use it with smeared gauge field if one ensures that
the admissibility is kept intact under the smearing process such as the
gradient flow.

In this work, we have treated $B_p$ as a flat background gauge field, but we can
perform its path integral to obtain the $SU(N)/\mathbb{Z}_N$ Yang--Mills theory.
Here, we can add the discrete theta angle as the local counter term, so there
are $N$ distinct $SU(N)/\mathbb{Z}_N$ Yang-Mills theory as clarified
in~Ref.~\cite{Aharony:2013hda}. In the continuum analysis, it is suggested that
they have dyonic line operators as gauge-invariant line operators, and their
electric charge is specified by the choice of the discrete theta parameter. In
the lattice formulation, we need to violate the flatness condition for~$B_p$
along the dyonic line. It would be an interesting future study to uncover
various topological phenomena, such as the Witten effect~\cite{Witten:1979ey},
on the lattice. As we have defined the admissible gauge fields only for $SU(N)$
link variables coupled with the flat $2$-form gauge fields in this paper, we
would need further generalizations for studying those phenomena.

It would also be interesting to extend this work to the case with fundamental
matter fields. In such cases, the flavor symmetry and the gauge redundancy has
the common center, and we may have a 't~Hooft anomaly between the projective 
nature of the flavor symmetry and the $\theta$ periodicity~\cite{Shimizu:2017asf,Gaiotto:2017tne,Tanizaki:2017qhf}. To obtain
such an anomaly, the background $2$-form gauge field plays an important role
even though the $1$-form symmetry is not present, so we expect that our result
is useful to give the fully lattice regularized derivation of those 't~Hooft
anomalies.

\acknowledgments

The authors appreciate the conference ``Lattice and continuum field theories 2022'' (YITP-W-22-02) at Yukawa Institute for Theoretical Physics (YITP) in the last July. 
This work was partially supported by Japan Society for the Promotion of Science
(JSPS) Grant-in-Aid for Scientific Research Grant Numbers JP21J30003 (O.M.),
JP20H01903 (H.S.), JP22H01218, and JP20K22350 (Y.T.).
The work of Y.T. was supported by Center for Gravitational Physics and Quantum
Information (CGPQI) at YITP.


\appendix

\section{Fractional power of special unitary matrices}
\label{sec:fractional_power}

In construction of transition functions, we need to take a fractional
power~$U^y$ with~$0\le y\le1$ for special unitary matrices~$U\in SU(N)$. In
this section, we give the definition including its domain and describe its
properties that are necessary in the discussion in the main text.

Let $\mathfrak{su}(N)$ be the Lie algebra of~$SU(N)$, which is given
by~$N\times N$ Hermitian traceless matrices. We have the exponential map,
\begin{equation}
   \exp:\mathfrak{su}(N)\to SU(N),\qquad X\mapsto\exp(\im X),
\end{equation}
and this map is surjective, which means that any $U\in SU(N)$ can be written
as~$U=\exp(\im X)$ for some $X\in\mathfrak{su}(N)$. Thus, we are tempted to
define $U^y=\exp(\im yX)$, but this is not well defined as the exponential map
is not injective. To circumvent this issue, we restrict the domain of the
exponential map to the following subset,
\begin{equation}
   \calD
   =\left\{
   X\in\mathfrak{su}(N)\bigm|\text{eigenvalues of $X\in(-\pi,\pi)$}
   \right\},
\label{eq:domain_exp}
\end{equation}
that is, when $X$ is diagonalized as~$\diag(\theta_1,\dotsc,\theta_N)$
with~$\theta_1+\dotsb+\theta_N=0$ then $X\in\calD$ if and only if the
eigenvalues satisfy $|\theta_k|<\pi$ for all $k=1$, \dots, $N$. The
exponential map is injective on~$\calD$, and thus $U^y$ can be defined as
\begin{equation}
   U^y=\exp(\im yX)
\label{eq:fractional_power}
\end{equation} 
for $U=\exp(\im X)\in\exp(\im\calD)\subset SU(N)$. When
$U\in SU(N)\setminus\exp(\im\calD)$, we do not define $U^y$.

In order to get some insight, let us discuss what kinds of elements in~$SU(N)$
are excluded from the definition of~$U^y$. We can immediately see that the
nontrivial center elements, such as $\rme^{\frac{2\pi\im}{N}}\bm{1}\in SU(N)$,
are excluded from the domain since
\begin{equation}
   \rme^{\frac{2\pi\im}{N}}\bm{1}
   =\exp\left[\frac{2\pi\im}{N}\diag(1,\dotsc,1,1-N)\right],
\end{equation}
which shows that $\theta_N=\frac{2\pi}{N}(1-N)$ does not fit the condition. Our
definition also excludes 
\begin{equation}
   \diag(-1,-1,1,\dotsc,1)
   =\exp[\diag(\im\pi,-\im\pi,0,\dotsc,0)],
\end{equation}
which is a nontrivial center element for a subgroup $SU(2)\subset SU(N)$.
Especially when $N=2$, this is the element treated as the exceptional
configuration in~Ref.~\cite{Luscher:1981zq}.

Let $V\in SU(N)$. Since the set of eigenvalues are not affected by the
conjugate operation, $VXV^{-1}\in\calD$ if~$X\in\calD$. This shows that,
for~$U\in\exp(\im\calD)\subset SU(N)$,
\begin{equation}
   \left(VUV^{-1}\right)^y=VU^yV^{-1}.
\label{eq:conjugate}
\end{equation}
This is the key property of~$U^y$ for the discussions in the main text.

\section{Explicit formulas}

In this appendix, we give the concrete expressions for the $1$-form gauge
invariant links in the complete axial gauge, $\Tilde{u}^{n}_{xy}$, and the
$1$-form covariant transition function, $\Tilde{v}_{n,\mu}(x)$, at the corner
of~$f(n,\mu)$. 

\subsection{Formulas for $\Tilde{u}^{n}_{xy}$}
\label{sec:formula_axial_gauge}

\subsubsection*{For $\mu=4$.}

\begin{align}
   \Tilde{u}_{s_2s_7}^{n,n-\Hat{4}}
   &=\rme^{-\frac{2\pi\im}{N}B_{23}(n)}u_{s_2s_7}^{n,n-\Hat{4}},
\notag\\
   \Tilde{u}_{s_1s_5}^{n,n-\Hat{4}}
   &=\rme^{-\frac{2\pi\im}{N}B_{13}(n)}u_{s_1s_5}^{n,n-\Hat{4}},
\notag\\
   \Tilde{u}_{s_5s_4}^{n,n-\Hat{4}}
   &=\rme^{-\frac{2\pi\im}{N}B_{12}(n+\Hat{3})}u_{s_5s_4}^{n,n-\Hat{4}},
\notag\\
   \Tilde{u}_{s_6s_4}^{n,n-\Hat{4}}
   &=\rme^{-\frac{2\pi\im}{N}B_{23}(n)}\rme^{-\frac{2\pi\im}{N}B_{13}(n+\Hat{2})}
   u_{s_6s_4}^{n,n-\Hat{4}},
\notag\\
   \Tilde{u}_{s_1s_6}^{n,n-\Hat{4}}
   &=e^{-\frac{2\pi i}{N}B_{12}(n)}u_{s_1s_6}^{n,n-\Hat{4}}.
\end{align}
Others are trivial: 
$\Tilde{u}_{s_0s_3}^{n,n-\Hat{4}}=
\Tilde{u}_{s_3s_7}^{n,n-\Hat{4}}=
\Tilde{u}_{s_0s_2}^{n,n-\Hat{4}}=
\Tilde{u}_{s_3s_5}^{n,n-\Hat{4}}=
\Tilde{u}_{s_0s_1}^{n,n-\Hat{4}}=
\Tilde{u}_{s_7s_4}^{n,n-\Hat{4}}=
\Tilde{u}_{s_2s_6}^{n,n-\Hat{4}}=\bm{1}$. 

\subsubsection*{For $\mu=3$.}
\begin{align}
   \Tilde{u}_{s_2s_7}^n
   &=\rme^{-\frac{2\pi\im}{N}B_{24}(n)}u_{s_2s_7}^n,
\notag\\
   \Tilde{u}_{s_1s_5}^n
   &=\rme^{-\frac{2\pi\im}{N}B_{14}(n)}u_{s_1s_5}^n,
\notag\\
   \Tilde{u}_{s_5s_4}^n
   &=\rme^{-\frac{2\pi\im}{N}B_{12}(n+\Hat{4})}u_{s_5s_4}^n,
\notag\\
   \Tilde{u}_{s_6s_4}^n
   &=\rme^{-\frac{2\pi\im}{N}B_{24}(n)}\rme^{-\frac{2\pi\im}{N}B_{14}(n+\Hat{2})}
   u_{s_6s_4}^n,
\notag\\
   \Tilde{u}_{s_1s_6}^n
   &=\rme^{-\frac{2\pi\im}{N}B_{12}(n)}u_{s_1s_6}^n.
\end{align}
Others are trivial:
$\Tilde{u}_{s_0s_3}^n=
\Tilde{u}_{s_3s_7}^n=
\Tilde{u}_{s_0s_2}^n=
\Tilde{u}_{s_3s_5}^n=
\Tilde{u}_{s_0s_1}^n=
\Tilde{u}_{s_7s_4}^n=
\Tilde{u}_{s_2s_6}^n=\bm{1}$. 

\begin{align}
   \Tilde{u}_{s_0s_3}^{n-\Hat{3}}
   &=\rme^{-\frac{2\pi\im}{N}B_{34}(n-\Hat{3})}u_{s_0s_3}^{n-\Hat{3}},
\notag\\
   \Tilde{u}_{s_2s_7}^{n-\Hat{3}}
   &=\rme^{-\frac{2\pi\im}{N}B_{34}(n-\Hat{3})}\rme^{-\frac{2\pi\im}{N}B_{24}(n)}
   u_{s_2s_7}^{n-\Hat{3}},
\notag\\
   \Tilde{u}_{s_1s_5}^{n-\Hat{3}}
   &=\rme^{-\frac{2\pi\im}{N}B_{34}(n-\Hat{3})}\rme^{-\frac{2\pi\im}{N}B_{14}(n)}
  u_{s_1s_5}^{n-\Hat{3}},
\notag\\
   \Tilde{u}_{s_5s_4}^{n-\Hat{3}}
   &=\rme^{-\frac{2\pi\im}{N}B_{12}(n+\Hat{4})}u_{s_5s_4}^{n-\Hat{3}},
\notag\\
   \Tilde{u}_{s_6s_4}^{n-\Hat{3}}
   &=\rme^{-\frac{2\pi\im}{N}B_{34}(n-\Hat{3})}
   \rme^{-\frac{2\pi\im}{N}B_{24}(n)}\rme^{-\frac{2\pi\im}{N}B_{14}(n+\Hat{2})}
   u_{s_6s_4}^{n-\Hat{3}},
\notag\\
   \Tilde{u}_{s_1s_6}^{n-\Hat{3}}
   &=\rme^{-\frac{2\pi\im}{N}B_{12}(n)}u_{s_1s_6}^{n-\Hat{3}}.
\end{align}
Others are trivial:
$\Tilde{u}_{s_3s_7}^{n-\Hat{3}}=
\Tilde{u}_{s_0s_2}^{n-\Hat{3}}=
\Tilde{u}_{s_3s_5}^{n-\Hat{3}}=
\Tilde{u}_{s_0s_1}^{n-\Hat{3}}=
\Tilde{u}_{s_7s_4}^{n-\Hat{3}}=
\Tilde{u}_{s_2s_6}^{n-\Hat{3}}=\bm{1}$. 

\subsubsection*{For $\mu=2$.}

\begin{align}
   \Tilde{u}_{s_2s_7}^n
   &=\rme^{-\frac{2\pi\im}{N}B_{34}(n)}u_{s_2s_7}^n,
\notag\\
   \Tilde{u}_{s_1s_5}^n
   &=\rme^{-\frac{2\pi\im}{N}B_{14}(n)}u_{s_1s_5}^n,
\notag\\
   \Tilde{u}_{s_5s_4}^n
   &=\rme^{-\frac{2\pi\im}{N}B_{13}(n+\Hat{4})}u_{s_5s_4}^n,
\notag\\
   \Tilde{u}_{s_6s_4}^n
   &=\rme^{-\frac{2\pi\im}{N}B_{34}(n)}\rme^{-\frac{2\pi\im}{N}B_{14}(n+\Hat{3})}
   u_{s_6s_4}^n,
\notag\\
   \Tilde{u}_{s_1s_6}^n
   &=\rme^{-\frac{2\pi\im}{N}B_{13}(n)}u_{s_1s_6}^n.
\end{align}
Others are trivial:
$\Tilde{u}_{s_0s_3}^n=
\Tilde{u}_{s_3s_7}^n=
\Tilde{u}_{s_0s_2}^n=
\Tilde{u}_{s_3s_5}^n=
\Tilde{u}_{s_0s_1}^n=
\Tilde{u}_{s_7s_4}^n=
\Tilde{u}_{s_2s_6}^n=\bm{1}$. 

\begin{align}
   \Tilde{u}_{s_0s_3}^{n-\Hat{2}}
   &=\rme^{-\frac{2\pi\im}{N}B_{24}(n-\Hat{2})}u_{s_0s_3}^{n-\Hat{2}},
\notag\\
   \Tilde{u}_{s_3s_7}^{n-\Hat{2}}
   &=\rme^{-\frac{2\pi\im}{N}B_{23}(n-\Hat{2}+\Hat{4})}u_{s_3s_7}^{n-\Hat{2}},
\notag\\
   \Tilde{u}_{s_2s_7}^{n-\Hat{2}}
   &=\rme^{-\frac{2\pi\im}{N}B_{34}(n-\Hat{2})}\rme^{-\frac{2\pi\im}{N}B_{24}(n-\Hat{2}+\Hat{3})}
   u_{s_2s_7}^{n-\Hat{2}},
\notag\\
   \Tilde{u}_{s_0s_2}^{n-\Hat{2}}
   &=\rme^{-\frac{2\pi\im}{N}B_{23}(n-\Hat{2})}u_{s_0s_2}^{n-\Hat{2}},
\notag\\
   \Tilde{u}_{s_1s_5}^{n-\Hat{2}}
   &=\rme^{-\frac{2\pi\im}{N}B_{24}(n-\Hat{2})}\rme^{-\frac{2\pi\im}{N}B_{14}(n)}
   u_{s_1s_5}^{n-\Hat{2}},
\notag\\
   \Tilde{u}_{s_5s_4}^{n-\Hat{2}}
   &=\rme^{-\frac{2\pi\im}{N}B_{23}(n-\Hat{2}+\Hat{4})}\rme^{-\frac{2\pi\im}{N}B_{13}(n+\Hat{4})}
   u_{s_5s_4}^{n-\Hat{2}},
\notag\\
   \Tilde{u}_{s_6s_4}^{n-\Hat{2}}
   &=\rme^{-\frac{2\pi\im}{N}B_{34}(n-\Hat{2})}\rme^{-\frac{2\pi\im}{N}B_{24}(n-\Hat{2}+\Hat{3})}
   \rme^{-\frac{2\pi\im}{N}B_{14}(n+\Hat{3})}
   u_{s_6s_4}^{n-\Hat{2}},
\notag\\
   \Tilde{u}_{s_1s_6}^{n-\Hat{2}}
   &=\rme^{-\frac{2\pi\im}{N}B_{23}(n-\Hat{2})}e^{-\frac{2\pi i}{N}B_{13}(n)}
   u_{s_1s_6}^{n-\Hat{2}}.
\end{align}
Others are trivial:
$\Tilde{u}_{s_3s_5}^{n-\Hat{2}}=
\Tilde{u}_{s_0s_1}^{n-\Hat{2}}=
\Tilde{u}_{s_7s_4}^{n-\Hat{2}}=
\Tilde{u}_{s_2s_6}^{n-\Hat{2}}=\bm{1}$.

\subsubsection*{For $\mu=1$.}
\begin{align}
   \Tilde{u}_{s_2s_7}^n
   &=\rme^{-\frac{2\pi\im}{N}B_{34}(n)}u_{s_2s_7}^n,
\notag\\
   \Tilde{u}_{s_1s_5}^n
   &=\rme^{-\frac{2\pi\im}{N}B_{24}(n)}u_{s_1s_5}^n,
\notag\\
   \Tilde{u}_{s_5s_4}^n
   &=\rme^{-\frac{2\pi\im}{N}B_{23}(n+\Hat{4})}u_{s_5s_4}^n,
\notag\\
   \Tilde{u}_{s_6s_4}^n
   &=\rme^{-\frac{2\pi\im}{N}B_{34}(n)}\rme^{-\frac{2\pi\im}{N}B_{24}(n+\Hat{3})}
   u_{s_6s_4}^n,
\notag\\
   \Tilde{u}_{s_1s_6}^n
   &=\rme^{-\frac{2\pi\im}{N}B_{23}(n)}u_{s_1s_6}^n. 
\end{align}
Others are trivial:
$\Tilde{u}_{s_0s_3}^n=
\Tilde{u}_{s_3s_7}^n=
\Tilde{u}_{s_0s_2}^n=
\Tilde{u}_{s_3s_5}^n=
\Tilde{u}_{s_0s_1}^n=
\Tilde{u}_{s_7s_4}^n=
\Tilde{u}_{s_2s_6}^n=\bm{1}$.

\begin{align}
   \Tilde{u}_{s_0s_3}^{n-\Hat{1}}
   &=\rme^{-\frac{2\pi\im}{N}B_{14}(n-\Hat{1})}u_{s_0s_3}^{n-\Hat{1}},
\notag\\
   \Tilde{u}_{s_3s_7}^{n-\Hat{1}}
   &=\rme^{-\frac{2\pi\im}{N}B_{13}(n-\Hat{1}+\Hat{4})}u_{s_3s_7}^{n-\Hat{1}},
\notag\\
   \Tilde{u}_{s_2s_7}^{n-\Hat{1}}
   &=\rme^{-\frac{2\pi\im}{N}B_{34}(n-\Hat{1})}\rme^{-\frac{2\pi\im}{N}B_{14}(n-\Hat{1}+\Hat{3})}
   u_{s_2s_7}^{n-\Hat{1}},
\notag\\
   \Tilde{u}_{s_0s_2}^{n-\Hat{1}}
   &=\rme^{-\frac{2\pi\im}{N}B_{13}(n-\Hat{1})}u_{s_0s_2}^{n-\Hat{1}},
\notag\\
   \Tilde{u}_{s_1s_5}^{n-\Hat{1}}
   &=\rme^{-\frac{2\pi\im}{N}B_{24}(n-\Hat{1})}\rme^{-\frac{2\pi\im}{N}B_{14}(n-\Hat{1}+\Hat{2})}
   u_{s_1s_5}^{n-\Hat{1}},
\notag\\
   \Tilde{u}_{s_5s_4}^{n-\Hat{1}}
   &=\rme^{-\frac{2\pi\im}{N}B_{23}(n-\Hat{1}+\Hat{4})}
   \rme^{-\frac{2\pi\im}{N}B_{13}(n-\Hat{1}+\Hat{2}+\Hat{4})}
   u_{s_5s_4}^{n-\Hat{1}},
\notag\\
   \Tilde{u}_{s_6s_4}^{n-\Hat{1}}
   &=\rme^{-\frac{2\pi\im}{N}B_{34}(n-\Hat{1})} \rme^{-\frac{2\pi\im}{N}B_{24}(n-\Hat{1}+\Hat{3})}
   \rme^{-\frac{2\pi\im}{N}B_{14}(n-\Hat{1}+\Hat{2}+\Hat{3})}
   u_{s_6s_4}^{n-\Hat{1}},
\notag\\
   \Tilde{u}_{s_1s_6}^{n-\Hat{1}}
   &=\rme^{-\frac{2\pi\im}{N}B_{23}(n-\Hat{1})}\rme^{-\frac{2\pi\im}{N}B_{13}(n-\Hat{1}+\Hat{2})}
   u_{s_1s_6}^{n-\Hat{1}},
\notag\\
   \Tilde{u}_{s_3s_5}^{n-\Hat{1}}
   &=\rme^{-\frac{2\pi\im}{N}B_{12}(n-\Hat{1}+\Hat{4})}
   u_{s_3s_5}^{n-\Hat{1}},
\notag\\
   \Tilde{u}_{s_0s_1}^{n-\Hat{1}}
   &=\rme^{-\frac{2\pi\im}{N}B_{12}(n-\Hat{1})}u_{s_0s_1}^{n-\Hat{1}},
\notag\\
   \Tilde{u}_{s_7s_4}^{n-\Hat{1}}
   &=\rme^{-\frac{2\pi\im}{N}B_{12}(n-\Hat{1}+\Hat{3}+\Hat{4})}u_{s_7s_4}^{n-\Hat{1}},
\notag\\
   \Tilde{u}_{s_2s_6}^{n-\Hat{1}}
   &=\rme^{-\frac{2\pi\im}{N}B_{12}(n-\Hat{1}+\Hat{3})}u_{s_2s_6}^{n-\Hat{1}}.
\end{align}

\subsection{Formulas for $\Tilde{v}_{n,\mu}(x)$}
\label{sec:transition_1formcov}

We find that the following expressions fulfill the requirements in the main text:

\subsubsection*{For $\mu=4$}
\begin{equation}
   \Tilde{v}_{n,4}(x)=v_{n,4}(x).
\end{equation}

\subsubsection*{For $\mu=3$}
\begin{equation}
   \Tilde{v}_{n,3}(x)
   =\begin{cases}
   v_{n,3}(x),&\text{for $x=s_0$, $s_1$, $s_2$, $s_6$},\\
   \rme^{\frac{2\pi\im}{N}B_{34}(n-\Hat{3})}
   v_{n,3}(x),&\text{for $x=s_3$, $s_4$, $s_5$, $s_7$}.\\
   \end{cases}
\end{equation}

\subsubsection*{For $\mu=2$}
\begin{equation}
   \Tilde{v}_{n,2}(x)
   =\begin{cases}
   v_{n,2}(x),&\text{for $x=s_0$, $s_1$},\\
   \rme^{\frac{2\pi\im}{N}B_{23}(n-\Hat{2})}
   v_{n,2}(x),&\text{for $x=s_2$, $s_6$}.\\
   \rme^{\frac{2\pi\im}{N}B_{24}(n-\Hat{2})}
   v_{n,2}(x),&\text{for $x=s_3$, $s_5$}.\\
   \rme^{\frac{2\pi\im}{N}B_{24}(n-\Hat{2})}
   \rme^{\frac{2\pi\im}{N}B_{23}(n-\Hat{2}+\Hat{4})}
   v_{n,2}(x),&\text{for $x=s_4$, $s_7$}.\\
   \end{cases}
\end{equation}

\subsubsection*{For $\mu=1$}
\begin{equation}
   \Tilde{v}_{n,1}(x)
   =\begin{cases}
   v_{n,1}(x),&\text{for $x=s_0$},\\
   \rme^{\frac{2\pi\im}{N}B_{12}(n-\Hat{1})}
   v_{n,1}(x),&\text{for $x=s_1$},\\
   \rme^{\frac{2\pi\im}{N}B_{13}(n-\Hat{1})}
   v_{n,1}(x),&\text{for $x=s_2$},\\
   \rme^{\frac{2\pi\im}{N}B_{14}(n-\Hat{1})}
   v_{n,1}(x),&\text{for $x=s_3$},\\
   \rme^{\frac{2\pi\im}{N}B_{14}(n-\Hat{1})}
   \rme^{\frac{2\pi\im}{N}B_{13}(n-\Hat{1}+\Hat{4})}
   \rme^{\frac{2\pi\im}{N}B_{12}(n-\Hat{1}+\Hat{3}+\Hat{4})}
   v_{n,1}(x),&\text{for $x=s_4$},\\
   \rme^{\frac{2\pi\im}{N}B_{14}(n-\Hat{1})}
   \rme^{\frac{2\pi\im}{N}B_{12}(n-\Hat{1}+\Hat{4})}
   v_{n,1}(x),&\text{for $x=s_5$},\\
   \rme^{\frac{2\pi\im}{N}B_{13}(n-\Hat{1})}
   \rme^{\frac{2\pi\im}{N}B_{12}(n-\Hat{1}+\Hat{3})}
   v_{n,1}(x),&\text{for $x=s_6$},\\
   \rme^{\frac{2\pi\im}{N}B_{14}(n-\Hat{1})}
   \rme^{\frac{2\pi\im}{N}B_{13}(n-\Hat{1}+\Hat{4})}
   v_{n,1}(x),&\text{for $x=s_7$}.\\
   \end{cases}
\end{equation}

\section{Bound on $\ve$ for the admissibility condition}
\label{sec:bound_epsilon}

Let $V$ be the vector space with the norm $\|\cdot\|$. For a linear
map~$A:V\to V$, its matrix norm is defined by
\begin{equation}
   \|A\|
   :=\sup_{\psi\in V,\|\psi\|=1}\|A\psi\|
   =\sqrt{\lambda_{\mathrm{max}}(A^\dagger A)},
\label{eq:matrix_norm}
\end{equation}
where $\lambda_{\mathrm{max}}$ is the maximum eigenvalue. For $U^y$
with~$0\le y\le1$ being well-defined, $U\in\exp(\im\calD)$ as discussed
in~Eq.~\eqref{eq:domain_exp}. We can characterize $U\in\exp(\im\calD)$ using
the matrix norm in the following way:
\begin{equation}
   U\in\exp(\im\calD)
   \Leftrightarrow
   \begin{aligned}
   &\text{There exists a map $t\in[0,1]\mapsto U(t)\in SU(N)$ such that}
   \\
   &\text{$U(0)=\bm{1}$, $U(1)=U$, and
   $\|U(t)-\bm{1}\|<2$ for all $t\in[0,1]$.}       
   \end{aligned}
\label{eq:DefByNorm}
\end{equation}
We can show $\Rightarrow$ by taking $U(t)=U^t$. To show $\Leftarrow$, let us
diagonalize $U\in SU(N)\setminus\exp(\im\calD)$, then their eigenvalues
$\rme^{\im\theta_i}$ with the choice of the branch $\theta_1+\dotsb+\theta_N=0$
should have $\max_i|\theta_i|\ge\pi$. Then, there is no way to circumvent
$\max_i|\theta_i(t_*)|=\pi$ for some $t=t_*$ when we connect $\bm{1}$ and~$U$,
at which $\|U(t_*)-\bm{1}\|=2$. 

Let $U_1$ and~$U_2$ be unitary matrices. We pick up $\psi\in V$
with~$\|\psi\|=1$, and then
\begin{align}
   \|(U_1U_2-\bm{1})\psi\|
   &=\|(U_1-\bm{1})U_2\psi+(U_2-\bm{1})\psi\|
\notag\\
   &\le\|(U_1-\bm{1})U_2\psi\|+\|(U_2-\bm{1})\psi\|
\notag\\
   &\le\|U_1-\bm{1}\|+\|U_2-\bm{1}\|.
\end{align}
By taking $\sup_\psi$ of the both sides, we obtain the useful inequality,
\begin{equation}
   \|U_1U_2\dotsb U_n-\bm{1}\|\le\sum_{i=1}^n\|U_i-\bm{1}\|.
\label{eq:norm_inequality}
\end{equation}
Combined with Eqs.~\eqref{eq:DefByNorm} and~\eqref{eq:norm_inequality}, we find
that for $U_1$, \dots, $U_n\in\exp(\im\calD)$,
\begin{equation}
   \|U_i-\bm{1}\|<\frac{2}{n}\Rightarrow
   U_1\dotsb U_n\in\exp(\im\calD).
\label{eq:ineq_admissibile}
\end{equation}
To prove it, let us set $U(t)=U_1^t\dotsb U_n^t$, then $U(0)=\bm{1}$
and~$U(1)=U_1\dotsb U_n$. Since $\|U(t)-\bm{1}\|\le\sum_i\|U_i^t-\bm{1}\|\le
\sum_i\|U_i-\bm{1}\|<\frac{2}{n}\cdot n=2$, we obtain the result
from~Eq.~\eqref{eq:DefByNorm}.

When we have products of $M$ plaquettes $\Tilde{U}_{p_i}$ ($i=1$, \dots, $M$),
we can take its fractional power $(\Tilde{U}_{p_1}\dotsb\Tilde{U}_{p_M})^y$ if
we impose the admissibility condition $\|\Tilde{U}_{p_i}-\bm{1}\|<\ve$ with
some $\ve\le2/M$ thanks to~Eq.~\eqref{eq:ineq_admissibile}. The transition
functions $\Tilde{v}_{n,\mu}(x)$ are defined via
$\Tilde{S}_{n,\mu}^m(x_\alpha,x_\beta,x_\gamma)$ given
in~Eq.~\eqref{eq:interpolater}, and the most severe bound for~$\ve$ is obtained
by the number of plaquettes appearing in~$\Tilde{l}_{n,\mu}^m(x_\beta,x_\gamma)$
for~$m=n-\hat{1}$. The number of plaquettes contained in~$\Tilde{u}_{xy}^{m}$
can be counted by the number of~$B_{\mu\nu}$ given
in~Appendix~\ref{sec:formula_axial_gauge}. We further note that $\Tilde{f}$,
$\Tilde{g}$, $\Tilde{h}$, $\Tilde{k}$ contain the combinations such as
\begin{equation}
   \Tilde{u}_{s_0s_3}^{n-\Hat{1}}
   \Tilde{u}_{s_3s_7}^{n-\Hat{1}}
   \Tilde{u}_{s_7s_2}^{n-\Hat{1}}
   \Tilde{u}_{s_2s_0}^{n-\Hat{1}}, 
\end{equation}
and this is nothing but the plaquette
$p=\langle s_0\to s_3\to s_7\to s_2\to s_0\rangle$ in the complete axial gauge.
Thus, we can simply count it as one plaquette, and we then find that there are
$M=27$ plaquettes in the $\Tilde{l}_{n,\mu}^{n-\hat{1}}$. This shows that we can
take the admissibility condition given with
\begin{equation}
   \ve<\frac{2}{27}\simeq0.074.
\end{equation}
This bound is larger than that for the overlap Dirac operator
in~Ref.~\cite{Hernandez:1998et},
\begin{equation}
    \ve_{\mathrm{overlap}}<\frac{1}{30}\simeq0.033,
\end{equation}
but smaller than L\"uscher's assertion in~Ref.~\cite{Luscher:1981zq}:
$\ve<\sqrt{0.03}\simeq0.173$. We point out that
$\Tilde{f}$, $\Tilde{g}$, $\Tilde{k}$, $\Tilde{l}$, $\Tilde{S}$ take the form
of~$U_1^{-y}(U_1U_2U_3^{-1}U_4^{-1})^yU_4U_3^y=
U_1^{1-y}U_2(U_2^{-1}U_1^{-1}U_4U_3)^{1-y}U_3^{y-1}$. We can obtain larger
bounds for~$\ve$ by estimating their norm as a function of~$y_\beta$ 
and~$y_\gamma$ as it improves the above discussion, but we do not pursue it
here.

\bibliographystyle{utphys}
\bibliography{./QFT.bib,./refs.bib}
\end{document}